\documentclass{PoS}

\title{Observation of the extremely bright flare of the FSRQ 3C279 with H.E.S.S. II}

\ShortTitle{3C 279 with H.E.S.S. II}

\author{\speaker{C. Romoli} $^a$, M. Zacharias$^{b}$, M. Meyer$^{c}$, F. Ait Benkhali$^d$, A. Jacholkowska$^e$, A. Wierzcholska$^f$, F. Jankowsky$^g$, J.P. Lenain$^e$ for the H.E.S.S. Collaboration\\
\llap{$^a$} Dublin Institute for Advanced Studies, 31 Fitzwilliam Place, Dublin 2, Ireland \\
\llap{$^b$} Centre for Space Research, North-West University, Potchefstroom 2520, South Africa \\
\llap{$^c$} Oskar Klein Centre, Department of Physics, Stockholm University, Albanova University Center, SE-10691 Stockholm,Sweden \\
\llap{$^d$} Max-Planck-Institut f\"ur Kernphysik, P.O. Box 103980, D 69029 Heidelberg, Germany \\
\llap{$^e$} Sorbonne Universit\'es, UPMC Universit\'e Paris 06, Universit\'e Paris Diderot, Sorbonne Paris Cit\'e, CNRS, Laboratoire de Physique Nucl\'eaire et de Hautes Energies (LPNHE), 4 placeJussieu, F-75252, Paris Cedex 5, France\\
\llap{$^f$} Instytut Fizyki J\c{a}drowej PAN, ul. Radzikowskiego 152, 31-342 Krak\'ow, Poland\\
\llap{$^g$} Landessternwarte, Universit\"at Heidelberg, K\"onigstuhl, D69117 Heidelberg, Germany\\

E-mail: \email{romolic@cp.dias.ie}}

\abstract{In June 2015, the Flat Spectrum Radio Quasar 3C 279 underwent an extremely bright gamma-ray flare, with an increase of the flux above 100 MeV by a factor 10 in less than 1 day, revealing an intrinsic variability timescale of 2 minutes as detected by the Fermi-LAT.
We present results of target of opportunity observations with the H.E.S.S. experiment on this source over the nights around the peak of the outburst.
The H.E.S.S. data were analysed with mono and stereo chains. Thanks to the extreme brightness of the source at GeV energies, it was possible to obtain data from Fermi-LAT, strictly simultaneous to the H.E.S.S. observation. Simultaneous and quasi-simultaneous observations at optical and X-ray energies were gathered to reconstruct the multiwavelength spectrum helping to constrain theoretical models describing the flare.
The H.E.S.S. observation during the second night, using H.E.S.S. II MONO data, lead to a clear detection of the source in about 3 hours of live-time. The H.E.S.S. results were also used to derive limits on the Quantum Gravity scale under the assumption of Lorentz Invariance Violation. Furthermore, since FSRQs possess intense optical photon fields surrounding the central region near the black hole, the VHE data allows constraints on the location of the emitting region to be derived in order that internal absorption be avoided.
The detection of VHE emission from the powerful flare of the FSRQ 3C 279 by H.E.S.S. II can provide unique insights into the physical properties of this class of blazar, thanks in part to the presence of simultaneous and quasi-simultaneous datasets at other wavelengths. Due to the high redshift of the source (z=0.54), it was also possible to derive strong constraints on the Quantum Gravity mass scale.}

\FullConference{35th International Cosmic Ray Conference $-$ ICRC2017-\\
		10-20 July, 2017\\
		Bexco, Busan, Korea}

\begin{document}

\section{Introduction}

The source 3C~279 is one of the most famous gamma-ray emitters in the GeV sky, it is found at a redshift $z=0.536$ \cite{1965ApJ...142.1667L} and it is a highly variable source. It belongs to the class of the Flat Spectrum Radio Quasars (FSRQs), which are believed to be the version of the powerful FR~II radio galaxies in which the jet is pointed towards the observer \cite{1995PASP..107..803U}. The multiwavelength Spectral Energy Distribution (SED) is characterized by a double peaked structure with a first bump at infrared energies (associated to synchrotron emission of relativistic electrons) and a high energy peak centred between 100~MeV and few GeV. On the origin of this latter component there is still some debate. In leptonic models it is associated to inverse Compton emission of electrons up-scattering photons from the various ambient radiation fields, while in the hadronic scenarios, the high energy emission arises mostly from proton synchrotron (see e.g. \cite{1997AIPC..410..494S}). 

In the high energy range ($100 \textrm{ MeV} \leq E \leq 100 \textrm{ GeV}$), the source was clearly detected with EGRET (between the years 1991 and 2000) \cite{1999ApJS..123...79H} and subsequently by the \textit{Fermi}-LAT \cite{2015ApJS..218...23A}. Using ground based telescopes, 3C~279 was detected with the MAGIC telescope during outbursts in 2006 \cite{2008Sci...320.1752M} and 2007 \cite{2011A&A...530A...4A}, while other attempts to detect this source were unsuccessful \cite{2014A&A...567A..41A, 2016AJ....151..142A}.

In June 2015, the source underwent a massive flare at GeV energies, reaching its historical maximum in the high energy gamma-ray band, with a peak flux of $\left(3.6 \pm 0.2\right)\times10^{-5}$ ph/cm$^2$/s \cite{2016ApJ...824L..20A}, a factor $>20$ more than the preflare state. To enhance the statistical signal, the \textit{Fermi} satellite was operating in ``pointing mode" and this has allowed the possibility to detect minute-scale variability in the flux above 100~MeV \cite{2016ApJ...824L..20A}. Following the \textit{Fermi}-LAT observations of the increasing flux, announced on June, the 15$^{th}$ 2015 \cite{2015ATel.7633....1C}, the H.E.S.S. Collaboration Triggered a Target of Opportunity on the source.

\section{H.E.S.S. Observation}

The H.E.S.S. array is a hybrid system of Imaging Atmospheric Cherenkov Telescopes located in the Kohmas Highlands of Namibia at 1800 m of altitude. The array is composed by 4 telescopes with a reflector diameter of 12 m positioned at the vertexes of a square of 120 m side (the H.E.S.S. I array), while at the centre of the square, since 2012 it is operative a much larger ($\sim28$ m diameter) reflector named CT5, aimed to lower the energy threshold of the array below 100~GeV, entering the H.E.S.S. II phase of the telescope.

\begin{figure}[ht]
\centering
\includegraphics[width=0.75\textwidth]{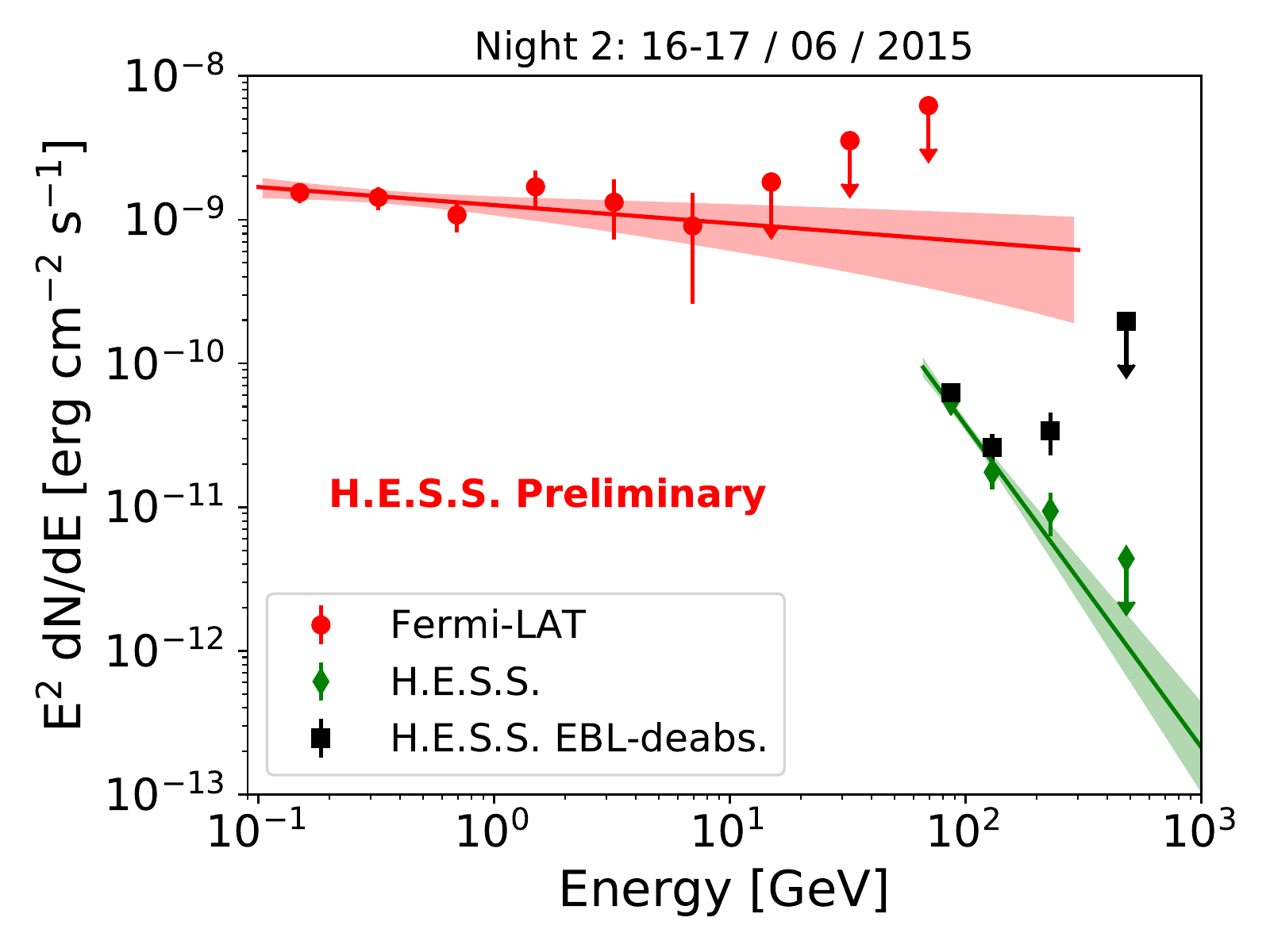}
\caption{Gamma ray SED of 3C~279 during the H.E.S.S. observation in the second night, between $16^{th}$ and $17^{th}$ of June 2015. In red the \textit{Fermi}-LAT data points with the best fit PL and the 1~$\sigma$ envelope. In green the H.E.S.S. II MONO data and in black the H.E.S.S. spectrum of the source corrected for the EBL absorption.}
\label{fig:hess_spectrum}
\end{figure}

The H.E.S.S. observations were performed around the peak of the flare detected by the \textit{Fermi}-LAT and were carried out for 5 consecutive nights starting from the 15$^{th}$-$16^{th}$ of June, but only during the second and fourth night it was possible to use CT5. The observations during this second night were crucial and, thanks to the combination of the high flux of the source and the low threshold allowed by CT5, 3C~279 was detected at 8.7 confidence level in 2.2 hours (acceptance corrected live-time) above an energy threshold of 66 GeV. The measured differential flux could be fitted with a power law function in the form $dN/dE = N_0\left(E/E_0\right)^{-\Gamma}$ and the parameters are reported in Table~\ref{tab:fitres} together with the simultaneous \textit{Fermi}-LAT results and the quasi simultaneous observations in X-rays performed by the \textit{Swift} satellite. In Figure~\ref{fig:hess_spectrum} is shown the SED from the H.E.S.S. and \textit{Fermi}-LAT data for the second night. In this plot is also present the spectrum de-absorbed for the effect of the Extragalactic Background Light (EBL) according to the model proposed by Franceschini et al. (2008) \cite{2008A&A...487..837F}.

Through the H.E.S.S. observations, for photon energies above 66~GeV, it was not possible to detect any significant flux variability.

\section{Multiwavelength data}

\begin{figure}
\centering
\includegraphics[width=0.45\textwidth]{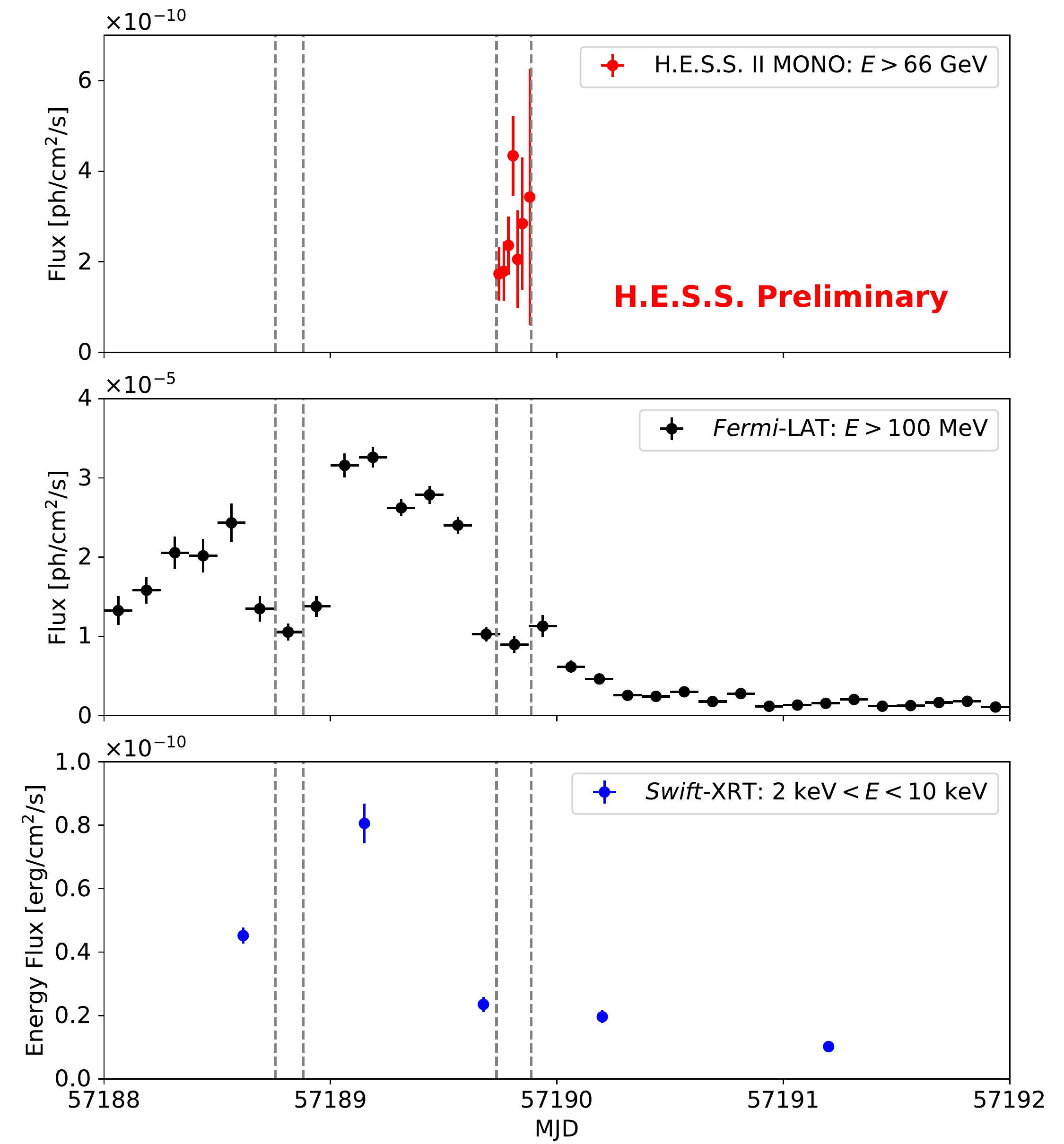}
\includegraphics[width=0.45\textwidth]{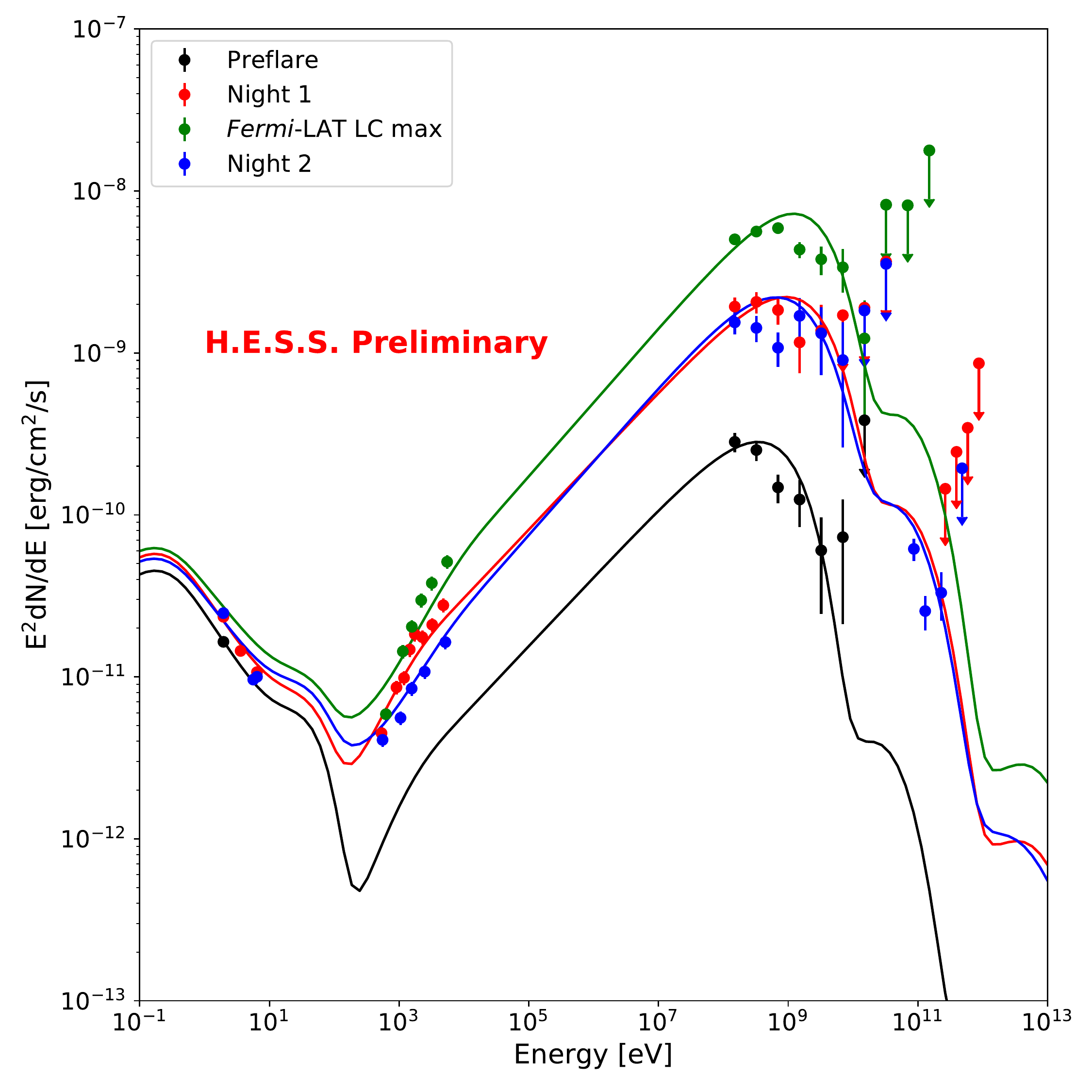}
\caption{ \textbf{Left:} Multiwavelength light curve from Swift-XRT (lower panel), Fermi-LAT (middle panel) and HESS (top panel). The plot highlights in particular the high variability of the X-rays and of the GeV gamma-ray band. The dashed lines indicate the time intervals of the H.E.S.S. observations. \textbf{Right:} Multiwavelength SED for the various periods in Table 1 with superimposed the fit with a hadronic model by (cfr. text for reference).}
\label{fig:mwl_lc}
\end{figure}

Due to the brightness of the high energy flare, several other telescopes, at different wavelengths observed the source. A multiwavelength light curve is shown in the left panel of Figure~\ref{fig:mwl_lc}.

The \textit{Fermi}-LAT data were analysed using the PASS8 IRFs using the latest version of the Science Tools software (v10.0.5). The data were extracted from a square region with a side of 30 degrees. A binned likelihood fit was implemented to retrieve the source parameters, once taken into account all the sources listed in the 3FGL catalogue and the diffuse and isotropic background made available by the \textit{Fermi} Collaboration. The data were binned in a light-curve with 3 hour bin size, shown in the second panel of Figure~\ref{fig:mwl_lc}. The X-ray data instead were analysed with the HEAsoft software (v.6.16). All events were cleaned, calibrated, and analysed in the energy range of 0.3-10 keV. The light curve flux points were calculated from the spectra of a single observation integrating between 2 and 10 keV. For the spectral fitting, XSPEC v. 12.8.2 with a single power-law model and Galactic hydrogen absorption fixed to Galactic value of $n_H
= 2.01 \cdot10^{20}\;{\rm cm}^{-2}$ were used.

The light curve of the flux above 100 MeV shows a flux at the peak 40 times brighter than the average emission reported in the 3FGL catalogue. In Figure~\ref{fig:mwl_lc}, it is possible to appreciate the magnitude of this flaring event, produced by an enhancement of the emission of the high energy peak of the SED. The data collected by {\it Swift}-XRT in the rising part of this spectral peak show indeed a similar trend with respect to the LAT data. The low energy part of the SED, sampled by Swift-UVOT and by the ATOM telescope, does not show any variability comparable to what is seen at X and gamma-ray energies.

Table~\ref{tab:fitres} shows the power law fit on the H.E.S.S., Fermi-LAT on the simultaneous time intervals of the H.E.S.S. observations (night 1 and 2), together with the fit results on the peak interval of the light curve shown in Figure~\ref{fig:mwl_lc} and a pre-flare phase between the 11$^{th}$ and the 14$^{th}$ of June (MJD 57184-57187). The table reports as well the spectral fit of the Swift-XRT pointing closest to the chosen Fermi-LAT intervals.

\begin{table}[ht]
\centering
\footnotesize
\caption{Power law fit on the \textit{Fermi}-LAT, H.E.S.S. and \textit{Swift}-XRT data during simultaneous (or quasi-simultaneous in X-ray) observations. During night 1, CT5 was not available and it was retrieved only an upper limit. The table reports also the Fermi-LAT fit on a preflare interval (MJD 57184-57187) and the peak of the GeV light curve. All the errors are statistical only.}
\begin{tabular}{lcccccc}
&\multicolumn{2}{c}{\textit{Swift}-XRT} & \multicolumn{2}{c}{\textit{Fermi}-LAT} & \multicolumn{2}{c}{H.E.S.S.}\\
\cline{2-3} \cline{4-5} \cline{6-7}
& Diff. flux       &                    & Diff. flux         &                  & Diff. flux        &                  \\
& ($E=1$ keV)      & $\Gamma_{XRT} $    & ($E=342$ MeV)      & $\Gamma_{LAT}$   & ($E=98$ GeV)      & $\Gamma_{HESS} $ \\                       
& [ph/cm$^2$/s/keV]&                    & [ph/cm$^2$/s/GeV]  &                  & [ph/cm$^2$/s/GeV] &                  \\
\hline
\hline
Night 1  & $\left( 5.3\pm0.3\right)\times 10^{-3}$ & $1.26 \pm 0.05$& $\left(9.2 \pm 0.9\right)\times 10^{-6}$ & $2.2 \pm 0.1$   & \multicolumn{2}{c}{N.A. (upper limit)}    \\[.2em]
Night 2  & $\left(3.5\pm0.2\right)\times 10^{-3}$ & $1.36 \pm 0.07$& $\left(7.7 \pm 0.8\right)\times 10^{-6}$ & $2.1 \pm 0.1$   &  $\left(2.5\pm0.2\right)\times 10^{-12}$  & $4.2 \pm 0.3$ \\[.2em]
\hline
Preflare & - & - &$\left(1.1 \pm 0.1\right)\times 10^{-6}$  &$2.3\pm 0.1$     & - & - \\[.2em]
Peak     & $\left( 7.3\pm0.4\right) \times 10^{-3}$ & $1.08 \pm 0.06$& $\left(27 \pm 1\right)\times 10^{-6}$    & $2.09 \pm 0.04$ & - & - \\
\hline
\end{tabular}
\label{tab:fitres}
\end{table}

\section{Discussion}

The high quality of the data retrieved during the observation campaign, especially in the high energy range makes possible a discussion on the characterization of the high energy emission. This flare detected by the \textit{Fermi}-LAT was the second brightest since the operation of the spacecraft and the presence of strictly simultaneous data with H.E.S.S. allows us an in depth study of this source.

Given the FSRQ nature of 3C~279, beside the EBL, also the internal absorption due to the dense photon field of the broad line region can be very relevant in suppressing the very high energy flux. Making the assumption that the unabsorbed spectrum is the one measured by the Fermi-LAT up to hundreds of GeV, the internal optical depth $\tau$ can be retrieved through the comparison between this extrapolated flux and the flux seen by H.E.S.S. corrected for the EBL effect as done in \cite{pks1510flareicrc}. Using the lowest flux in the H.E.S.S. de-absorbed points, this simple calculation provides $\tau=3.3\pm0.7_{{\rm stat}}$. Taking as a reference the BLR model by Finke (2016) \cite{2016ApJ...830...94F}, the emitting region would be situated in approximately at the Lyman$_{\alpha}$ radius distance ($R_{Ly_{\alpha}}=1.1 \cdot 10^{17}$ cm) from the central black hole\footnote{$\sim R_{Ly_{\alpha}}$ distance for a ring geometry, while $\sim3R_{Ly_{\alpha}}$ for a shell model of the BLR \cite{2016ApJ...830...94F}}. The assumption of an intrinsic curvature in the spectrum of 3C~279, would move the emitting region towards larger distances from the central engine. 

To characterize the multiwavelength emission of the source during the flaring event, a leptonic model was not be able to obtain a reasonable fit of the SED due to the lack of strong variability in correspondence of the ``synchrotron peak" of the broad-band SED. A hadronic model (from Diltz et al. (2015) \cite{2015ApJ...802..133D}, shown in right panel of Figure~\ref{fig:mwl_lc}) was instead able to describe the more general characteristics of the flare, even though the flat spectra measured by the \textit{Fermi}-LAT are problematic to match in this framework. The final parameters in this hadronic context would require an initial injection of particles to mimic the first part of the flare followed by a subsequent harder injection of protons to explain the peak of the GeV light-curve. In this model the emission at GeV energy is directly due to the proton synchrotron emission, while the flux at higher energy is related to a muon synchrotron component.

A search for Lorentz Invariance Violation (LIV) effects was performed with 3C 279 flare data and lead to results on the Quantum Gravity (QG) energy scale. Thanks to the high redshift and the important number of events in a large energy range above 50 GeV, the source is a good candidate for this type of studies. The study of energy dependent time-delays within a deterministic scenario and with a Likelihood method was performed following \cite{2011APh....34..738H}. Two cases were considered: a linear and a quadratic dependence of the speed of light on the photon energy. From the absence of significant delay in the photon arrival time, limits on the QG energy scale were computed. The one-sided 95\% CL QG limits, including systematic uncertainties, for the sub-luminal and supra-luminal case are reported in Table~\ref{tab:liv}.

\begin{table}[h]
\centering
\caption{The 95\% limits on the QG scale derived from the 3C~279 observations by H.E.S.S. for a linear and a quadratic dependence on the photon speed with the energy, in the sub-luminal and super-luminal case.}
\begin{tabular}{lcc}
\hline
                & Sub-luminal              & Supra-luminal             \\
\hline
Linear dep.     & $1.7 \cdot 10^{17}$ GeV  & $3.8 \cdot 10^{17}$ GeV   \\
Quadratic dep.  & $2.0 \cdot 10^{10}$ GeV  & $3.7\cdot10^{10}$ GeV     \\
\hline
\end{tabular}
\label{tab:liv}
\end{table}

Even if less constraining than those obtained from studies with the PKS~2155$-$304 2006 flare \cite{2011APh....34..738H}, which were obtained with much higher statistics, the  3C~279 flare results with this very high redshift are extremely important as they confirm previous results with other AGNs and match the limits obtained with GRBs \cite{2013PhRvD..87l2001V}.

\section{Conclusions}

The FSRQ 3C~279 was detected at its historical maximum in the GeV band by the \textit{Fermi}-LAT in the June 2015. The target of opportunity that followed has allowed the H.E.S.S. telescope to detect the source using a H.E.S.S. II MONO configuration with a significance of $8.7\,\sigma$ in 2.2 hours of effective live-time with no significant flux variability. The reconstructed spectrum was a steep one with a photon index $\Gamma_{VHE}=4.2\pm0.3$. Considering the flare in a multi-wavelength view, this event involved mostly the high energy peak of the broadband SED, with no strong variability in the low energy peak, as visible from Figure~\ref{fig:mwl_lc}. 

Once the EBL effect has been subtracted from the H.E.S.S. data, the joint spectrum from \textit{Fermi}-LAT and H.E.S.S. can be used to determine the distance of the emitting region from the central engine. In a conservative assumption that does not require any intrinsic curvature of the spectrum, the emitting region would sit at least at a distance of $10^{17}$ cm from the central engine. When applying a hadronic model to the multiwavelength datasets needs two distinct injection events with the Fermi data explained by proton synchrotron emission and the component seen by HESS as the result of muon synchrotron. A leptonic model was not able to describe satisfactory the data. Finally, the H.E.S.S. detection with CT5 of this high red-shift source, it was possible to set limits on the quantum gravity scale through upper limits on Lorentz Invariance Violation that matched what was previously achieved with GRBs and other AGNs.

\section*{Acknowledgements}
The support of the Namibian authorities and of the University of Namibia in facilitating the construction and operation of H.E.S.S. is gratefully acknowledged, as is the support by the German Ministry for Education and Research (BMBF), the Max Planck Society, the German Research Foundation (DFG), the Alexander von Humboldt Foundation, the Deutsche Forschungsgemeinschaft, the French Ministry for Research, the CNRS-IN2P3 and the Astroparticle Interdisciplinary Programme of the CNRS, the U.K. Science and Technology Facilities Council (STFC), the IPNP of the Charles University, the Czech Science Foundation, the Polish National Science Centre, the South African Department of Science and Technology and National Research Foundation, the University of Namibia, the National Commission on Research, Science \& Technology of Namibia (NCRST), the Innsbruck University, the Austrian Science Fund (FWF), and the Austrian Federal Ministry for Science, Research and Economy, the University of Adelaide and the Australian Research Council, the Japan Society for the Promotion of Science and by the University of Amsterdam.
We appreciate the excellent work of the technical support staff in Berlin, Durham, Hamburg, Heidelberg, Palaiseau, Paris, Saclay, and in Namibia in the construction and operation of the equipment. This work benefited from services provided by the H.E.S.S. Virtual Organisation, supported by the national resource providers of the EGI Federation.


\begin{thebibliography}{99}
\bibitem{1965ApJ...142.1667L} Lynds, C.~R., Stockton, A.~N., \& Livingston, W.~C.\ 1965, \textit{ApJ}, 142, 1667 
\bibitem{1995PASP..107..803U} Urry, C.~M., \& Padovani, P.\ 1995, \textit{Publ. Astron. Soc. Pac.}, 107, 803 
\bibitem{1997AIPC..410..494S} Sikora, M.\ 1997, Proceedings of the Fourth Compton Symposium, 410, 494 
\bibitem{1999ApJS..123...79H} Hartman, R.~C., Bertsch, D.~L., Bloom, S.~D., et al.\ 1999, \textit{ApJS}, 123, 79 
\bibitem{2015ApJS..218...23A} Acero, F., Ackermann, M., Ajello, M., et al.\ 2015, \textit{ApJS}, 218, 23 
\bibitem{2008Sci...320.1752M} MAGIC Collaboration, Albert, J., Aliu, E., et al.\ 2008, Science, 320, 1752 
\bibitem{2011A&A...530A...4A} Aleksi{\'c}, J., Antonelli, L.~A., Antoranz, P., et al.\ 2011, \textit{A\&A}, 530, A4 
\bibitem{2014A&A...567A..41A} Aleksi{\'c}, J., Ansoldi, S., Antonelli, L.~A., et al.\ 2014, \textit{A\&A}, 567, A41 
\bibitem{2016AJ....151..142A} Archambault, S., Archer, A., Benbow, W., et al.\ 2016, \textit{AJ}, 151, 142 
\bibitem{2016ApJ...824L..20A} Ackermann, M., Anantua, R., Asano, K., et al.\ 2016, \textit{ApJL}, 824, L20 
\bibitem{2015ATel.7633....1C} Cutini, S.\ 2015, The Astronomer's Telegram, 7633,  
\bibitem{2008A&A...487..837F} Franceschini, A., Rodighiero, G., \& Vaccari, M.\ 2008, \textit{A\&A}, 487, 837 
\bibitem{pks1510flareicrc} Zacharias, M., Sitarek, J., Prester, D.~D., et al.\ 2017, \textit{PoS (ICRC2017)}655
\bibitem{2016ApJ...830...94F} Finke, J.~D.\ 2016, \textit{ApJ}, 830, 94 
\bibitem{2015ApJ...802..133D} Diltz, C., B{\"o}ttcher, M., \& Fossati, G.\ 2015, \textit{ApJ}, 802, 133 
\bibitem{2011APh....34..738H} H.E.S.S.~Collaboration, Abramowski, A., Acero, F., et al.\ 2011, Astroparticle Physics, 34, 738 
\bibitem{2013PhRvD..87l2001V} Vasileiou, V., Jacholkowska, A., Piron, F., et al.\ 2013, \textit{Phys. Rev. D}, 87, 122001 
\end{thebibliography}
\end{document}